\documentclass[manuscript]{acmart}

\usepackage{graphicx}
\usepackage[caption=false]{subfig}
\usepackage{subcaption}
\usepackage{hyperref}

\hypersetup{
    colorlinks=true,
    linkcolor=blue,
    filecolor=magenta,      
    urlcolor=magenta,
    pdftitle={Overleaf Example},
    pdfpagemode=FullScreen,
    }

\AtBeginDocument{%
  \providecommand\BibTeX{{%
    \normalfont B\kern-0.5em{\scshape i\kern-0.25em b}\kern-0.8em\TeX}}}

\setcopyright{acmlicensed}
\copyrightyear{2024}
\acmYear{2024}
\acmDOI{10.1145/3631700.3664869}

\acmConference[UMAP '24]{The 32nd ACM Conference on User Modeling, Adaptation and Personalization}{July 01--04,
  2024}{Cagliari, Sardinia, Italy}

\acmBooktitle{UMAP '24: ACM Conference on User Modeling, Adaptation and Personalization,
 July 01--04, 2024, Cagliari, Sardinia, Italy} 
\acmISBN{978-1-4503-XXXX-X/18/06}

\begin{document}

\title{Beyond Static Calibration: The Impact of User Preference Dynamics on Calibrated Recommendation}

\author{Kun Lin}
\affiliation{%
   \institution{DePaul University}
   \city{Chicago}
   \country{USA}}
\email{klin13@depaul.edu}

\author{Masoud Mansoury}
\affiliation{%
   \institution{Delft University of Technology}
   \city{Delft}
   \country{Netherlands}}
\email{m.mansoury@tudelft.nl}

\author{Farzad Eskandanian}
\affiliation{%
   \institution{DePaul University}
   \city{Chicago}
   \country{USA}}
\email{feskanda@depaul.edu}

\author{Milad Sabouri}
\affiliation{%
   \institution{DePaul University}
   \city{Chicago}
   \country{USA}}
\email{msabouri@depaul.edu}

\author{Bamshad Mobasher}
\affiliation{%
   \institution{DePaul University}
   \city{Chicago}
   \country{USA}}
\email{mobasher@cs.depaul.edu}

\renewcommand{\shortauthors}{Lin et al.}

\acmCodeLink{https://github.com/nicolelin13/DynamicCalibrationUMAP}
\acmDataLink{https://github.com/nicolelin13/DynamicCalibrationUMAP/Data}

\begin{abstract}
  Calibration in recommender systems is an important performance criterion that ensures consistency between the distribution of user preference categories and that of recommendations generated by the system. Standard methods for mitigating miscalibration typically assume that user preference profiles are static, and they measure calibration relative to the full history of user's interactions, including possibly outdated and stale preference categories. We conjecture that this approach can lead to recommendations that, while appearing calibrated, in fact, distort users' true preferences. In this paper, we conduct a preliminary investigation of recommendation calibration at a more granular level, taking into account evolving user preferences. By analyzing differently sized training time windows from the most recent interactions to the oldest, we identify the most relevant segment of user's preferences that optimizes the calibration metric. We perform an exploratory analysis with datasets from different domains with distinctive user-interaction characteristics. We demonstrate how the evolving nature of user preferences affects recommendation calibration, and how this effect is manifested differently depending on the characteristics of the data in a given domain. Datasets, codes, and more detailed experimental results are available at:  \href{https://github.com/nicolelin13/DynamicCalibrationUMAP}{https://github.com/nicolelin13/DynamicCalibrationUMAP}.
  
  
  
\end{abstract}



\maketitle

\section{Introduction}

Recommender systems learn from past user preferences in order to predict future user interests and provide users with personalized suggestions. Traditionally, these systems are evaluated using accuracy metrics like precision, recall, and normalized discounted cumulative gain (NDCG), measuring the relevancy of the recommended items delivered to the users. Optimizing recommender systems for accuracy, however, may result in an imbalance in the distribution of preference categories in recommendations: the recommendations may only reflect a user's dominant preferences in certain categories and ignore less dominant interests while still being considered accurate. This type of imbalance, in the long term, can lead to undesirable effects such as the "rabbit hole" effect \cite{o2015down,zhao2021rabbit} or filter bubbles \cite{nguyen2014exploring}.

To address this phenomenon, the notion of \textit{Calibration} in machine learning has been extended to the realm of recommender systems \cite{steck2018calibrated}. The calibration metric measures the degree to which a user’s prior preference are reflected in the recommendations generated by the system. For example, in a hypothetical movie recommender system, a user whose past preferences suggest 70\% interest in Action movies and 30\% in Drama movies, should ideally see the same genre ratios in her set of recommended movies. Calibration is typically measured as the correlation between the distribution of preference categories in a user's profile and that of preference categories in generated recommendations. 

To improve calibration (or reduce miscalibration), most existing approaches assume that users’ preferences distribution is fundamentally static. To measure calibrations, these approaches typically utilize the full history of items in users' profiles to identify preference distributions. This may result in presumably calibrated recommendations that do not truly reflect users' current preferences. The problem is that user preferences evolve over time. If a user's profile is collected over a long period of time, only certain segments of that profile may represent the relevant interests of users. 

For example, consider a user interacting with a book platform for a period of 10 years. For the first 7 years, the user was intensely interested in books in the spy-thriller category. But 3 years ago, her interests shifted primarily to historical novels and then, in the last two years, to science fiction. Using the standard approach to measuring calibration, a recommender that recommends items from all three categories based on the user's full history might be considered \textit{highly calibrated}. However, the fact is that the user may not perceive these recommendations as reflective of her current preferences. On the other hand, if a recommender only recommends items from the historical and science fiction categories (i.e., more current user's preferences), then the generated recommendations may be considered to be \textit{miscalibrated}. 

Recognizing that a key contributor to miscalibration may be the dynamic nature of users' preferences, we hypothesize that selectively utilizing the most relevant segments from users' profiles for training the recommendation model can result in recommendations that better represent users' current interests. Our work attempts to shift the focus from predominantly post-processing methods for re-calibrating recommendations to incorporating calibration directly into the training phase of the recommendation process, promising a more fundamental solution to the challenge of maintaining relevance in the face of evolving user preferences.

In this preliminary work, we conduct an exploratory analysis of calibration in recommendation systems at a more granular level, taking into account the dynamics of user preferences. We empirically identify relevant segments of users' profiles through a data-driven study of two different datasets with different user interaction characteristics. Our goal is to show how the existing approach of measuring calibration that is based on static user profiles can be misleading and fails to precisely measure the degree to which recommendations reflect current user interests. To do this, we propose a preprocessing simulation process that identifies the most representative segment of the users' profile by measuring changes in the calibration metric as user preferences shift over time. 

In this simulation process, we first split the users' profiles into a number of segments, with the size of segments determined empirically based on the frequency of user interactions with items. Then, from the most recent segments to the oldest, we iteratively combine the segments to obtain subsamples of the data. Finally, on each subsample, we train a recommendation model and measure recommendation calibration. This process returns the most relevant segment of a user's profile that results in the highest calibration with respect to the current user preference distribution. 

Our initial experiments with two datasets from distinctive domains show that our proposed simulation process enables measuring calibration more precisely based on users' changing preferences. We hope that this preliminary work will lead to more extensive studies of how the dynamics of user preferences in different domains should be reflected in the design of calibration mechanisms in recommender systems.

\section{Background and Related Work}

We denote $\mathcal{U}=\{u_1,...,u_N\}$ and $\mathcal{I}=\{i_1,...,i_M\}$ as the set of $N$ users and $M$ items in the system. We show a user $u$'s profile as $I_u$ which contains all interacted items by $u$. $R_u$ is the recommendation list delivered to $u$ which contains all items recommended to $u$. Each item is assigned one or more categories $\mathcal{C}=\{c_1,...,c_W\}$, with $W$ categories in total.

\subsection{Calibration Metric in Recommendation}

The user's preference distribution over item categories is represented as a numerical vector of size $W$ that shows the preference ratio of a user toward each item category. We denote this vector for a user $u$ as $\mathcal{P}_u=\{p(c|u) | c \in \mathcal{C}\}$ where $p(c|u)$ is defined as follows:
\begin{equation}
\vspace{-2pt}
    p(c|u)= \frac{\sum_{i \in I_u}{p(c|i)}}{|I_u|}
\end{equation}

\noindent where $p(c|i)$ is the fraction that category $c$ represents item $i$ (e.g., if $i$ is assigned $c_1$ and $c_2$, then $p(c_1|i)=0.5$ and $p(c_2|i)=0.5$). Analogously, we define category distribution on the recommendation list delivered to user $u$, denoted by $\mathcal{Q}_u\{q(c|u) | c \in \mathcal{C}\}$, where $q(c|u)$ is defined as follows:
\begin{equation}
\vspace{-2pt}
    q(c|u)= \frac{\sum_{i \in R_u}{p(c|i)}}{|R_u|}
\end{equation}

Given the category distribution over the user's profile and recommendation list, miscalibration is computed by measuring the distance between these two distributions. There are various distance measures to consider for this computation, but following \cite{steck2018calibrated}, we use Kullback-Leibler (KL) divergence in this paper:
\begin{equation}
\vspace{-2pt}
    MC(\mathcal{P},\mathcal{Q}) = KL(\mathcal{P}||\mathcal{Q}) = \sum_{c \in \mathcal{C}}{p(c|u) \log{\frac{p(c|u)}{q(c|u)}}}
\end{equation}

\subsection{Methods for Calibrated Recommendation} 

Steck \cite{steck2018calibrated} initially defined calibration  emphasizing its role in achieving personalization that mirrors a user's past behavior. This concept has evolved, intersecting significantly with fairness and diversity \cite{abdollahpouri2020connection, wang2023two}. 

Current calibration methods \cite{sonboli2020multiaspectfair, eskandanian2020stablematching, abdollahpouri2023calibrated}, such as those critiqued by \cite{naghiaei2022towards, nazari2022choice}, often separate personalization from calibration, leading to a potential reduction in recommendation accuracy. Additionally, these methods typically assume static user preferences, neglecting the dynamic nature of user interests, as highlighted by Zhao et al. \cite{zhao2021rabbit}. 

Building upon the insights from previous studies, this research proposes a novel approach to address miscalibration in recommender systems. Traditional calibration methods often rely on entire historical data, where older interactions may no longer reflect the user's latest preferences, leading to recommendations that are out of sync with current interests. By selectively utilizing the most relevant segments of users' profiles for training the model, our approach aims to incorporate calibration directly into the training phase, taking into account the evolving nature of user preferences.

\begin{figure}
    \centering
    \includegraphics[width=\linewidth]{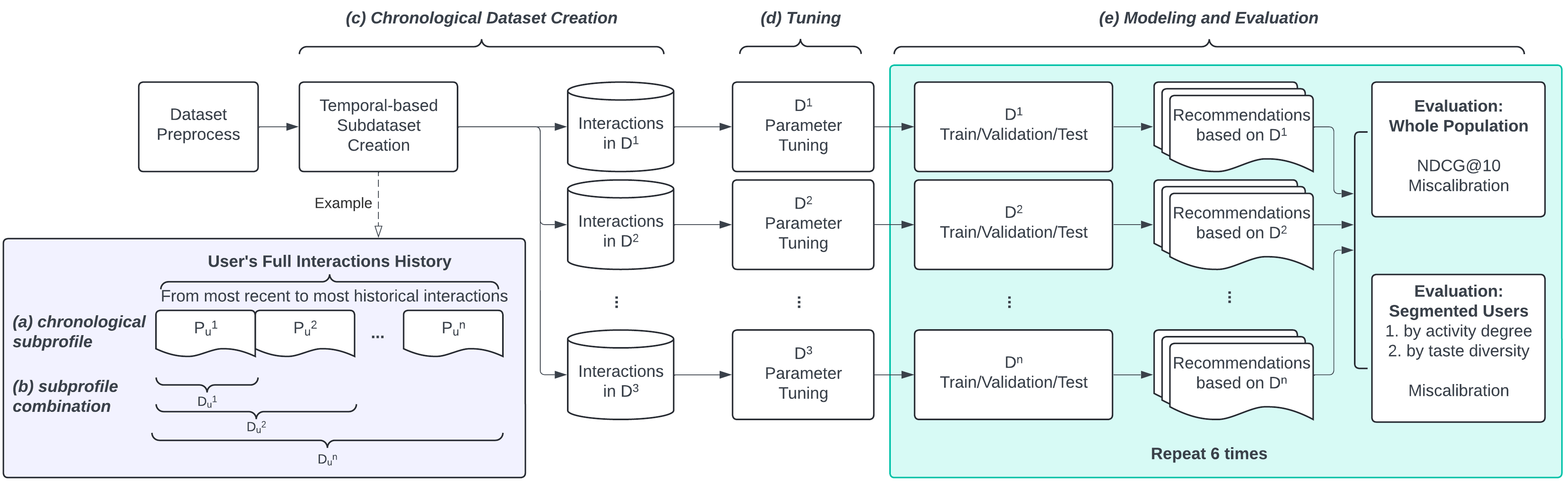}
    \vspace{-10pt}
    \caption{Experiment workflow for dynamic calibration}
    \label{fig:workflow}
    \vspace{-10pt}
\end{figure}

\section{Simulation Process for Dynamic Calibration}

In this section, we describe our pre-processing simulation process for identifying the most representative segments of users' profiles in training the recommendation model that would yield more calibrated recommendations. Figure \ref{fig:workflow} illustrates this simulation process as follows: 

\begin{enumerate}
    \item Given the user-item interaction matrix containing the profile of all users and their interacted items, we sort each user $u$'s profile $P_u$ chronologically in descending order and split $P_u$ into $n$ subprofiles as $\{P_u^1, P_u^2, ..., P_u^n\}$ where $P_u^1$ contains the most recent $u$'s interactions and $P_u^n$ contains the oldest interactions. This process is shown in part (a) of Figure \ref{fig:workflow}. The choice of $n$ (number of subprofiles) is a domain-specific parameter. The time \textit{window size} for creating the subprofiles should be set based on how frequently users interact with items. For example, in a short video streaming platform where users interact with items very frequently, a time window of shorter length, like daily, may be appropriate. However, in a book recommendation platform, a time window of a longer period, such as one or more years, might better capture users' evolving tastes. We discuss and analyze the choice of the time window and $n$ on different recommendation domains in the methodology section.

    \item As shown in part (b) of Figure \ref{fig:workflow}, we create samples of the dataset $D$ by iteratively and chronologically combining the subprofiles of users from different time windows as follows:
    \begin{equation}
    \vspace{-2pt}
        D^l = \{P_u^1 \cup P_u^2 \cup ... \cup P_u^l | \forall u \in \mathcal{U}\}, \;\;\;\; where \;\; l \leq n
    \end{equation}

    This will create samples of the dataset (i.e., $D^1, D^2, ..., D^n$) as shown in part (c) of Figure \ref{fig:workflow}.
    
    \item We iteratively pick each sample created in the previous step and evaluate the recommendation performance built on that sample. As shown in part (d) of Figure \ref{fig:workflow}, we first tune the recommendation model on each sample of the data for the best model. Then, in part (e) of Figure \ref{fig:workflow}, with the best model identified in the previous step, we evaluate the recommendation performance (i.e., miscalibration and accuracy) on each sample of the dataset. 
\end{enumerate}

When considering the different segments of the user's profile, the most recent segment of the users' profiles (i.e., $D^1$, $D^2$, ..., or $D^n$) that yields the lowest miscalibration is the one that is most representative of users' current preferences.

\section{Methodology}

\subsection{Datasets and Preprocessing}
We use two datasets from distinct domains. The short video platform is characterized by rapid user engagement and high interaction frequencies, where user preferences evolve quickly. For the book domain, users tend to transition between different book genres more gradually. To better investigate the calibration dynamics with the shifts in preference patterns, we preprocess datasets to focus on active users across time windows of different sizes. Detailed statistics of processed datasets are provided in Table \ref{tab:table_dataset}

KuaiRec dataset \cite{gao2022kuairec} is collected on the short video mobile app Kuaishou\footnote{https://www.kuaishou.com/cn} from July to September 2020. 
Each video is manually tagged with video categories. With watch\_ratio as the user's preference signal, we refer to its mean (0.9) as the threshold defining the user's interest. Positive interactions ($watch\_ratio\geq0.9$) are labeled as "1" with others as "0". Users in this dataset are very active. 86.4\% of users are active for 60 out of the total 63 days, while 99.9\% are active for more than 50 days. In our future work, we will include an analysis of the larger matrix from KuaiRec with higher sparsity which reflects a more realistic interaction scenario.

The book recommendation dataset \cite{wan2018item, wan2019fine} from GoodReads\footnote{https://mengtingwan.github.io/data/goodreads.html} includes user-book interactions. Each book has its assigned genres from all eight book genres. To reduce dataset size and sparsity from the original dataset we focused on interactions associated with more engaged users and frequently accessed items. The dataset was filtered to include only active users, defined as those engaging in 20 to 50 annual interactions from year 2010 to 2017. Subsequently, items with fewer than 1,000 interactions were excluded. The final step filtered in users with a consistent activity level of more than 4 interactions per year. 

\begin{table}
    \caption{Statistics of the Preprocessed Datasets.}
    \vspace{-10pt}
    \label{tab:table_dataset}
    \centering
    \begin{tabular}{lllllll}       
        \hline
        Dataset & \#Users & \#Items & Category type & \#Categories & \#Interactions & Sparsity\\
        \hline
        \hline
        KuaiRec   & 1,411 & 3,327 & content tag & 31 & 1,799,403 & 61.6\%\\        
        GoodReads & 865   & 1,662 & book genre  & 7  & 104,325   & 92.7\%\\
        \hline
    \end{tabular}
    \vspace{-9pt}
\end{table}

\subsection{Tuning Time Windows Representing User Preference Segments}
The time window selection over user profile is critical when creating users' subprofiles in Figure \ref{fig:workflow} part (a). In this exploratory work, we are interested in observing changes in calibration when extending the training time window sizes. Given that goal, we look for the smallest window size, leading to sufficient interactions for training and representing user preferences. Due to differences in user interaction frequency and evolving preferences across domains, time window sizes vary across domains. Users in the book reading domain have less frequent interactions and more stable preferences than users in the short video domain. After tuning the time window sizes for the two datasets (for KuaiRec, 1/7/14 days; for GoodReads, 3/6/12 months), we picked 1 day for KuaiRec and 0.5 years (6 months) for GoodReads.

\subsection{Recommendation Models}
Recommendations are generated by well-tuned Bayesian Personalized Ranking (BPR) \cite{rendle2012bpr}\footnote{Our full preliminary experiments include two algorithms: BPR and ItemKNN. Given the results between the two algorithms are similar, we focus on BPR in the main paper while making ItemKNN results available via \href{https://github.com/nicolelin13/DynamicCalibrationUMAP}{https://github.com/nicolelin13/DynamicCalibrationUMAP}.}, implemented using the library RecBole \cite{zhao2021recbole}. With NDCG@10 as objective, hyperparameters are tuned via an exhaustive search for learning rate ranging from 0.0001 to 0.01 and embedding size ranging between 64 to 128. Models for different time periods are independently tuned as shown in Figure \ref{fig:workflow} part (d). Also, we repeated the experiments six times to ensure the stability and accountability of the results. The results for analysis are based on the aggregated results from the repeated experiments.

\subsection{User Segment Analysis} 
Given that users' varied behavioral patterns in the same dataset may be related to different degrees of calibration across individuals, we extend our analysis of calibration dynamics to user segments with different characteristics. We explore two factors for segmenting users: user activity frequency (number of user interactions) and category-wise entropy in the user profile. We simply used a percentile-based approach, segmenting users into three groups based on each factor. A higher entropy indicates that user preferences are more evenly distributed across categories, while users with lower entropy have more intense preferences over a few specific categories. \cite{lin2020calibration} showed user profile entropy is an important user profile factor positively associated with the miscalibration of the received recommendation.

\section{Results and Discussion}

\subsection{Calibration Dynamics Due to Time Window Selections}
\begin{figure*}[t]
    \centering
    \vspace{-10pt}
    \subfloat[\centering KuaiRec]{\includegraphics[scale=0.165]{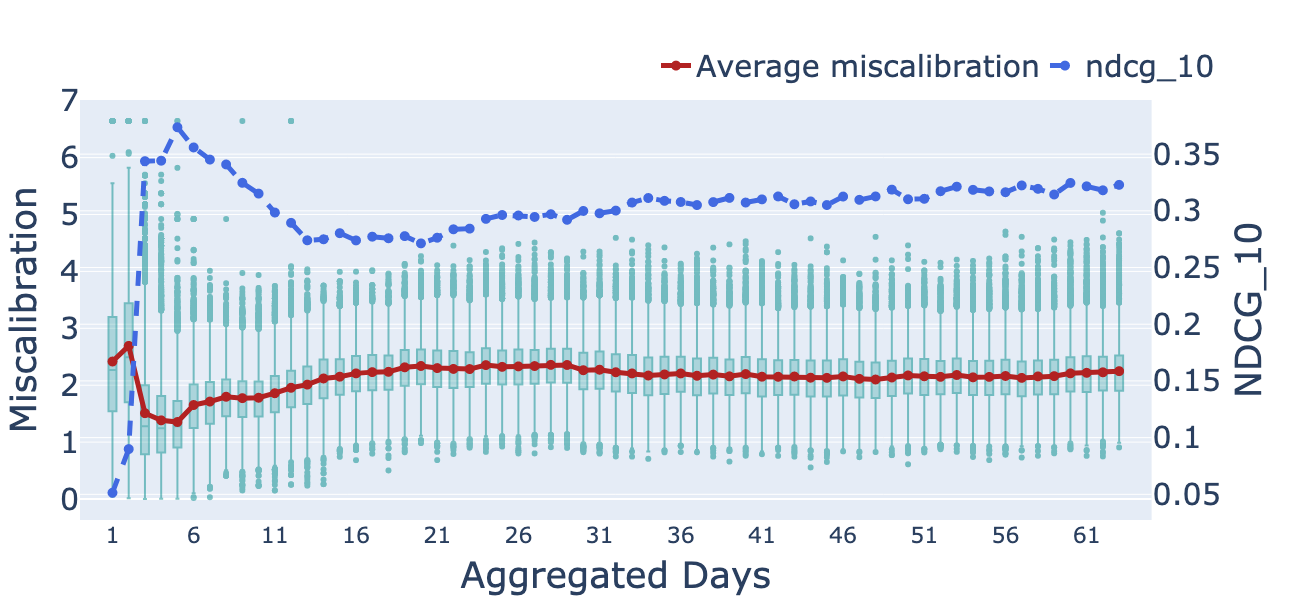}}
    \subfloat[\centering GoodReads]{\includegraphics[scale=0.165]{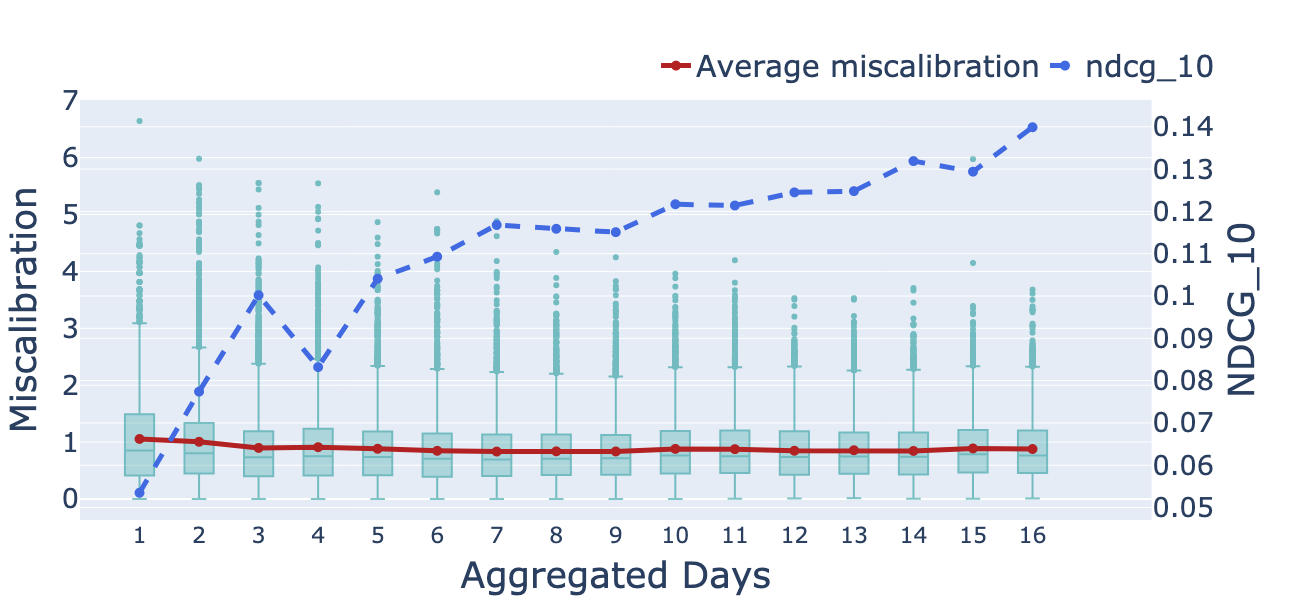}}
    \vspace{-8pt}
    \caption{Box plot of miscalibration distribution by time window}
    \label{fig:full}
    \Description{Box plot of miscalibration distribution by time window}
\end{figure*}

Fig. \ref{fig:full} depicts the miscalibration dynamics with the time window changing. Regardless of domains, we observe that miscalibration can vary based on the time window selections. In KuaiRec, the effect is more pronounced due to the highly dynamic nature of user preferences, with the optimal calibration achieved with a window size of 5 days. It is noteworthy that this window size also corresponds to the optimal NDCG values. In GoodReads, the calibration peaked at time window 7, indicating the most recent 3.5 years. In this domain, user preferences tend to be more stable over time, so there is less variance in calibration values. We note, however, that in this case, recommendation accuracy increases as larger profile segments are used in training.

\subsection{Calibration Dynamics Based on User Characteristics}

\begin{figure*}[t]
    \centering
    \subfloat[\centering Activity Degree]{\includegraphics[scale=0.165]{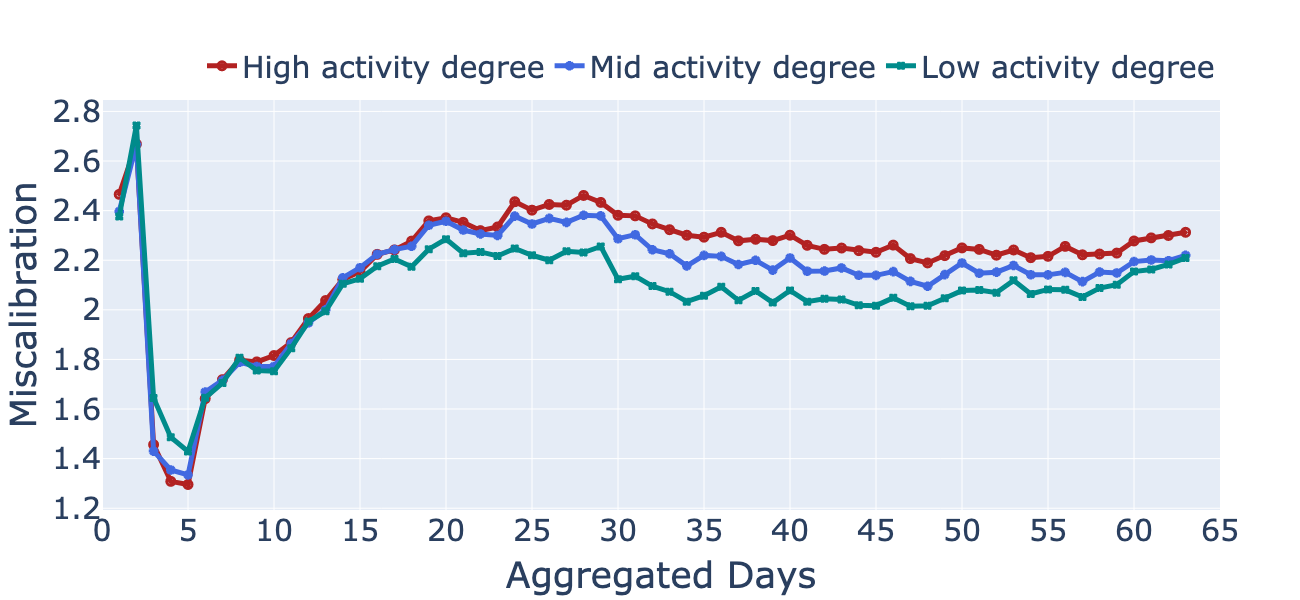}}
    \subfloat[\centering Profile Entropy]{\includegraphics[scale=0.165]{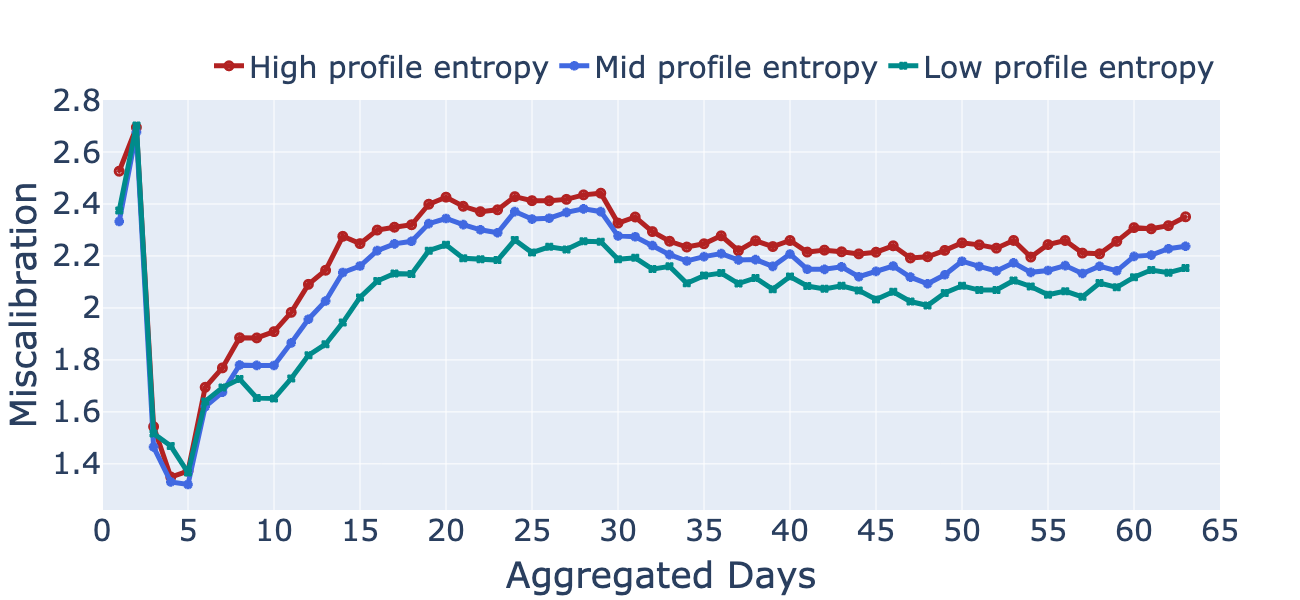}}
    \vspace{-8pt}
    \caption{Miscalibration by user segments in KuaiRec. The left is segmented by activity level, and the right is segmented by profile entropy.}
    \label{fig:kuairec_segment}
    \Description{Miscalibration by user segments in KuaiRec. The left is segmented by activity level, and the right is segmented by profile entropy.}
\end{figure*}

\begin{figure*}[t]
    \centering
    \subfloat[\centering Activity Degree]{\includegraphics[scale=0.165]{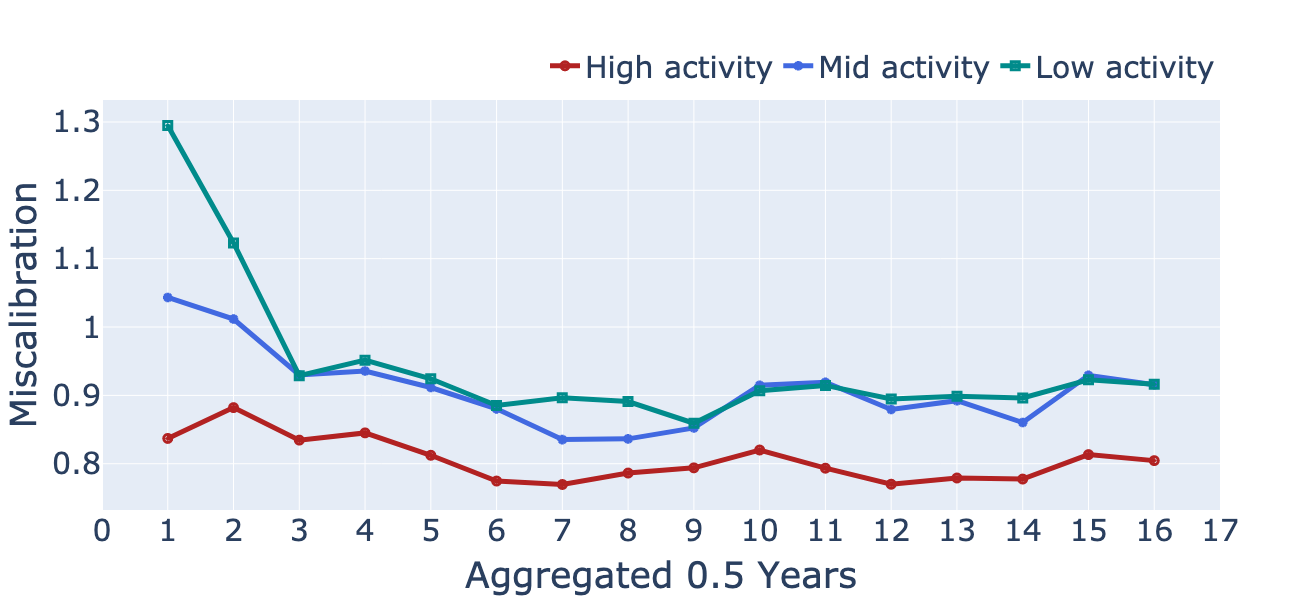}}
    \subfloat[\centering Profile Entropy]{\includegraphics[scale=0.165]{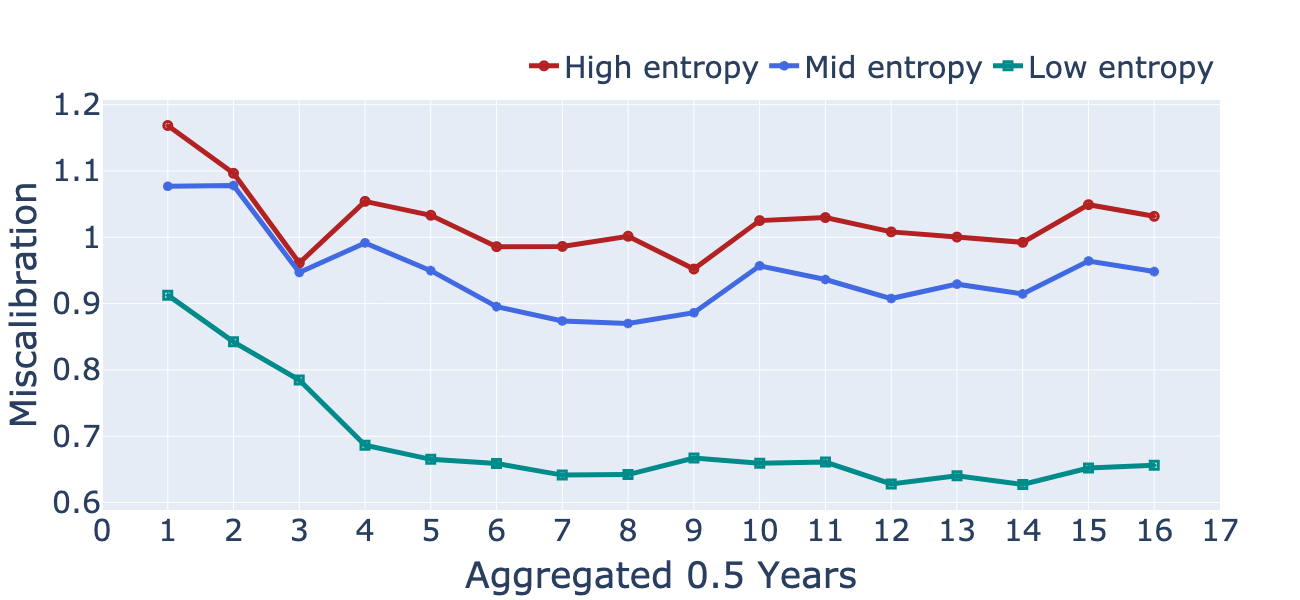}}
    \vspace{-8pt}
    \caption{Miscalibration by user segments in GoodReads. The left is segmented by activity level, and the right is segmented by profile entropy.}
    \label{fig:goodreads_segment}
    \Description{Miscalibration by user segments in GoodReads. The left is segmented by activity level, and the right is segmented by profile entropy.}
\end{figure*}

For our analysis of the impact of user segment characteristics on calibration dynamics, we empirically obtain the optimal time window for calibration as before, but we do so for each user segment instead of the full set of users.
Fig. \ref{fig:kuairec_segment}, depicts the changes in miscalibration as window size increases for the three user segments (with the left panel showing results for user segments based on interaction frequency and the right panel depicting results for user segments based on profile entropy). These results suggest that in this domain there are no significant variances in calibration across different user segments, especially when considering smaller time windows. This is likely due to the highly dynamic and unstable nature of user preferences. As before, we still see an optimal window size of 5 days for all user segments. 

Fig. \ref{fig:goodreads_segment} shows a very different picture for the GoodReads domain than that of KuaiRec. The optimal time window across user segments is not exactly consistent, but within the overall range that we saw for all users. Nevertheless, the results indicate that in this domain, it makes sense to consider different window sizes for distinct user segments to achieve optimal calibration. However, the more notable observation is that for user segments with a high activity level, a higher overall calibration is achieved regardless of window size. We conjecture that this is because users with higher activity in this domain (users who read books often) tend to have more stable profiles with little or no change in preference patterns over time. This is also the case for user segments with low profile entropy. These are users who tend to have very focused interests in a few genres, and so their preference patterns tend to be quite stable. 

Overall, these results support the conjecture that the degree of calibration in recommendation depends on the dynamics of user preferences, with more stable preference profiles resulting in a higher degree of calibration.

\section{Conclusion and Future Work}

In this exploratory analysis, we have shown that the calibration mechanism in recommender systems must take into account the dynamic nature of user preferences. Our simulation-based analysis shows that the degree of miscalibration decreases as the training time window size changes, focusing on the most relevant segments of the users' profiles. Our analysis also shows that user interaction characteristics associated with different domains lead to different optimal training window sizes to achieve better calibration. 

In future work, we extend this preliminary analysis to include comparison with post-processing calibration approaches, and also to explore trade-offs between dynamic calibration and other beyond accuracy metrics such as diversity and fairness. We will also conduct an analytical investigation on the impact of domain and user preference patterns on miscalibration dynamics. Currently, we have only considered two domains with a limited approach to segmenting usersand provided preliminary conclusions based on these observations. In future work, we will do a more comprehensive investigation involving additional data sets with distinct characteristics. We will also perform an analysis of how different classes of recommendation algorithms, including those based on session-based, context-aware, and sequence-aware recommendation models, perform using the dynamic approach to caliberation.

\pagebreak

\bibliographystyle{ACM-Reference-Format}
\bibliography{ref}

\end{document}